\documentclass[pss,fleqn,embeddedheads]{w-art}
\usepackage{times}
\usepackage{w-thm}
\usepackage[]{graphicx}
\newcommand\Yth{\ensuremath{\frac{\partial {Y_{\ell}}^{m}}{\partial \theta}}}
\newcommand\Yph{\ensuremath{\frac{\partial {Y_{\ell}}^{m}}{\partial \phi}}}
\newcommand\grad{\ensuremath{\mathbf{\nabla}}}
\newcommand\Ylm{\ensuremath{{Y_{\ell}}^{m}}}
\newcommand\er{\ensuremath{\hat{\mathbf{e}}_r}}
\newcommand\eth{\ensuremath{\hat{\mathbf{e}}_\theta}}
\newcommand\eph{\ensuremath{\hat{\mathbf{e}}_\phi}}
\newcommand\eone{\ensuremath{\textbf{\textit{e}}_1}}
\newcommand\etwo{\ensuremath{\textbf{\textit{e}}_2}}
\newcommand\ethree{\ensuremath{\textbf{\textit{e}}_3}}

\begin{document}

\title[Acoustic vibrations of an embedded nanoparticle]{The connection
between elastic scattering cross sections and acoustic vibrations of an
embedded nanoparticle}

\author{L. Saviot\footnote{Corresponding author: e-mail:
{\sf lucien.saviot@u-bourgogne.fr}}\inst{1}}
\address[\inst{1}]{Laboratoire de Recherche sur la R\'eactivit\'e
  des Solides, UMR 5613 CNRS -- Universit\'e de Bourgogne, Dijon, France}
\author{D. B. Murray\inst{2}}
\address[\inst{2}]{Department of Physics and Astronomy,
  Okanagan University College, Kelowna, British Columbia, Canada V1V 1V7}

\begin{abstract}
Arbitrary waves incident on a solid embedded nanoparticle
are studied.  The acoustic vibrational frequencies are shown
to correspond to the poles of the scattering cross section
in the complex frequency plane.
The location of the poles is unchanged
even
if the incident wave is nonplanar.
A second approach approximating the infinite matrix as a very large
shell surrounding the nanoparticle provides an
alternate way of predicting the mode frequencies.
The wave function of the vibration is also provided.
\end{abstract}

\maketitle

\section{Introduction}

In the last two decades, a significant number of works have been
devoted to the study of acoustic vibrations of spherical nanoparticles.
This has aided the understanding of the role of vibrations in the
performance of some optical devices (for example, in electronic
dephasing due to emission of phonons). These vibrations can be
experimentally investigated with different techniques including Raman
scattering\cite{Duval86}, photoluminescence\cite{Palinginis03}, and
femtosecond pump-probe experiments\cite{DelFattiJPC99}. Usually,
continuum elastic models have been used to describe these vibrations.
The original 1882 work by Lamb\cite{lamb1882} found the vibrational
frequencies of a free continuous isotropic elastic sphere. However in
real life, nanoparticles are seldom free but rather embedded in another
medium such as a glass matrix or liquid.

The problem of eigenmodes of vibrations of an embedded sphere has not
been extensively studied. However, a lot of work has been devoted
to the apparently unrelated problem of scattering of acoustic waves
by a sphere\cite{Brill87}. That is where Dubrovskiy \textit{et al.}
paper\cite{dubrovskiy81} has its starting point to calculate complex
natural frequencies of vibrations of an embedded sphere . Although
this method, referred to here as the complex frequency model (CFM),
is a very powerful tool, one might wonder about the significance
of the complex frequencies and the relevance of discrete natural
vibrations in a problem where one expects a continuum of vibrational
state. The goals of this paper are to explain the basics of this model
and to show its relevance by comparing it with the core shell model
(CSM)\cite{MurrayPRB04} which is a different approach using real valued
frequencies.

\section{Scattering of elastic waves and CFM}

The usual basis of unit vectors in spherical coordinates is
(\er, \eth, \eph). However, it is more convenient to express
results using a different set of vectors. In this paper, we
will use basic vectors
(\eone, \etwo, \ethree) defined as follows:
\begin{eqnarray}
	\eone   & = & \Ylm \er \\
	\etwo   & = & \frac1{\sin \theta} \Yph \, \eph + \Yth \, \eth \\
	\ethree & = & \frac1{\sin \theta} \Yph \, \eth - \Yth \, \eph
\end{eqnarray}
where \Ylm{} represents the spherical harmonics.

Vibrational modes of a sphere, whether free or
embedded, may be classified as ``torsional'' (TOR) or
``spheroidal'' (SPH).  Torsional modes are those for which
the displacement field has zero divergence.  All other modes
are called spheroidal.
Torsional modes have non-zero displacement along \ethree{} only.
Spheroidal modes have non-zero displacement along \eone{} and
\etwo{} only. For $\ell = 0$, we have $\etwo = \ethree = \mathbf{0}$.

\subsection{\bf Basic equations}
In order to clarify the meaning of the complex-valued
eigenfrequencies of an embedded elastic sphere,
the problem we want to investigate is the scattering of an
incident elastic wave in a medium (matrix) by a sphere made
of a different material than the matrix.  Three different
terms have to be considered: the incident
wave, the motion inside the sphere and the scattered wave.
All these waves have to satisfy the following wave equation
(assuming an $\exp(-i \omega t)$ time dependence) inside the
medium where they exist.
\begin{equation}
        (\lambda \!+\! 2 \mu) \mathbf{\nabla} ( \mathbf{\nabla} \!\!\cdot\! \textbf{\textit{u}} )
        - \mu \mathbf{\nabla} \!\!\times\! (\mathbf{\nabla} \!\!\times\! \textbf{\textit{u}}) + \rho \,
        \omega^{2} \textbf{\textit{u}} = 0
	\label{eqwave}
\end{equation}

$\textbf{\textit{u}}$ is the displacement, $\lambda$ and
$\mu$ the Lam\'e coefficients and $\rho$ the mass density.

Thus, they can be decomposed into the the following form:
\begin{equation}
        \textbf{\textit{u}} = \grad \Phi + \grad \!\!\times\! \left( \textbf{\textit{r}} \chi + \grad
        \!\!\times\! (\textbf{\textit{r}} \psi) \right)
\end{equation}
where the three potential fields are
\begin{eqnarray}
	\Phi    & = & \sum_{\ell=0,m} A_{\ell m} z_{\ell}(k_L r) \Ylm(\theta,\phi)\\
	\chi    & = & \sum_{\ell=1,m} B_{\ell m} z_{\ell}(k_T r) \Ylm(\theta,\phi)\\
	\psi    & = & \sum_{\ell=1,m} C_{\ell m} z_{\ell}(k_T r) \Ylm(\theta,\phi)
\end{eqnarray}
$z_\ell$ is the relevant spherical Bessel function of order $\ell$,
(first or second kind or Hankel)
$k_L$ and $k_T$ are longitudinal and transverse wavevectors
respectively.
\subsection{\bf Solutions}

Using the notations introduced by Eringen \textit{et al.}\cite{Eringen},
the displacement and the associated surface traction (surface force per
unit area) $\textbf{\textit{T}} =\sigma_{rr} \, \er + \sigma_{r\theta}
\, \eth + \sigma_{r\phi} \, \eph$ of the ($\ell$,$m$) component can be
written as follows:

\begin{equation}
	\begin{array}{c|c|c}
                potential & \textbf{\textit{u}} & \textbf{\textit{T}} \\
		\hline
		\Phi &
		U^z_1(\ell,k_L r) \, \eone + U^z_3(\ell,k_L r) \, \etwo &
		T^z_{11}(\ell,k_L r) \, \eone + T^z_{13}(\ell,k_L r) \, \etwo\\

		\chi &
		V^z_2(\ell,k_T r) \, \ethree &
		T^z_{42}(\ell,k_T r) \, \ethree\\

		\psi &
		V^z_1(\ell,k_T r) \, \eone + V^z_3(\ell,k_T r) \, \etwo &
		T^z_{41}(\ell,k_T r) \, \eone + T^z_{43}(\ell,k_T r) \, \etwo

	\end{array}
\end{equation}

The $z$ superscript has been added to recall the nature of
the spherical Bessel function to use.  Inside the sphere,
$z$ is the spherical Bessel function of the first kind $j$.
Outside the sphere, in the general case, two terms have to
be considered where $z$ is $j$ and $n$ (spherical Bessel
function of the second kind).  For the case of the outgoing
scattered wave, we'll use the spherical Hankel function of
the first kind $h^{(1)}_\ell(x)=j_\ell(x)+in_\ell(x)$.

\subsection{\bf Boundary conditions}

The boundary conditions are the usual continuities of
displacement and surface traction at the sphere-matrix
interface.  Because the boundary conditions should be
satisfied for $r=R$ (sphere radius) at all $\theta$ and
$\phi$, each equation of continuity breaks up into a
linear inhomogeneous system of six
equations
for each
$(\ell,m)$ the incident wave is decomposed onto. The
unknowns are the amplitudes of the vibration inside the
sphere and of the scattered wave.  The constants are the
amplitudes of the decomposition of the incident wave on
$(\ell,m)$.  Therefore the unknowns can be expressed as
the ratio of two $6 \times 6$ determinants.  The
determinant
equation in
the denominator is the one of the
homogeneous system, \textit{i. e.} for $\ell > 0$ and
$| m | \leq \ell$:

\begin{equation}
	D_{\ell} =
	\begin{array}{|cccccc|}
		U_1^j(\ell,\xi) &  U_3^j(\ell,\eta) & U_1^{h^{(1)}}(\ell,\xi) & U_3^{h^{(1)}}(\ell,\eta) & 0 & 0\\
		V_1^j(\ell,\xi) &  V_3^j(\ell,\eta) & V_1^{h^{(1)}}(\ell,\xi) & V_3^{h^{(1)}}(\ell,\eta) & 0 & 0\\
		\mu_p T_{11}^j(\ell,\xi)         & \mu_p T_{13}^j(\ell,\eta) &
		\mu_m T_{11}^{h^{(1)}}(\ell,\xi) & \mu_m T_{13}^{h^{(1)}}(\ell,\eta) & 0 & 0\\
		\mu_p T_{41}^j(\ell,\xi)         & \mu_p T_{43}^j(\ell,\eta) &
		\mu_m T_{41}^{h^{(1)}}(\ell,\xi) & \mu_m T_{43}^{h^{(1)}}(\ell,\eta) & 0 & 0\\
		0 & 0 & 0 & 0 &          V_2^j(\ell,\eta) &         V_2^{h^{(1)}}(\ell,\eta)\\
		0 & 0 & 0 & 0 & \mu_p T_{42}^j(\ell,\eta) & \mu_m T_{42}^{h^{(1)}}(\ell,\eta)
	\end{array}
\end{equation}
where $\xi$ = $k_L R$, $\eta$ = $k_T R$, $\mu_p$ and $\mu_m$ are the particle
and matrix mass densities and
the rows represent the continuity of the displacement along
$\textbf{\textit{u}} \!\cdot\! \eone$,
$\textbf{\textit{u}} \!\cdot\! \etwo$,
$\textbf{\textit{T}} \!\!\cdot\! \eone$,
$\textbf{\textit{T}} \!\!\cdot\! \etwo$,
$\textbf{\textit{u}} \!\cdot\! \ethree$ and
$\textbf{\textit{T}} \!\!\cdot\! \ethree$
from top to bottom.
The columns represent the coefficients for
$\Phi_{p}$,
$\psi_{p}$,
$\Phi_{m}$,
$\psi_{m}$,
$\chi_{p}$, and
$\chi_{m}$
where the $p$ and $m$ subscripts are for waves inside the particle and the
outgoing wave inside the matrix respectively.

Since the matrix is block diagonal this determinant is
easily decomposed as the product of
$\Delta_{\ell}$ ($4 \times 4$
determinant for SPH modes) and $\Gamma_{\ell}$
($2 \times 2$ determinant for TOR modes).

For $\ell$=$0$, both $\etwo$ and $\ethree$ equal $\mathbf{0}$
and the determinant is:
\begin{equation}
	D_0 = \Delta_0 =
	\begin{array}{|cc|}
		U_1^j(\ell=0,\xi) &       U_1^{h^{(1)}}(\ell=0,\xi) \\
		\mu_p T_{11}^j(\ell=0,\xi) & \mu_m T_{11}^{h^{(1)}}(\ell=0,\xi) 
	\end{array}
\end{equation}

These determinants do not depend on
the
azimuthal quantum number
$m$.  They are also independent of the nature of the
incident wave. This makes the results equally
applicable to a situation where the incident wave is a
spherical wave emanating from a source close by to the
scattering sphere.  This would be the case in a phononic
crystal, where elastic waves scatter sequentially from
a periodic array of scatterers.

\subsection{\bf CFM}
Up to now, only real valued $\omega$ have been considered.
The amplitudes of the wave inside the sphere and of the
outgoing wave have been shown to be the ratio of two
determinants with the same determinant
in
the denominator
(the determinant of the homogeneous linear system).  These
amplitudes can therefore become quite large when this
determinant is close to zero.  The physical system
corresponding to the system of homogeneous equations is the
one where there is no incident wave.  In this case, the only
solution is obtained when all the amplitudes are zero.
Therefore $D_{\ell}$ does not vanish for real values of $\omega$.
However, it can reach local minima. In order to predict the
position of these minima, we are going to locate
the complex roots of
the
matrix determinant $D_{\ell}$.
This is the heart of the complex frequency model introduced
by Dubrovskiy and Morochnik\cite{dubrovskiy81}. Because of
the $\exp(-i \omega t)$ dependence of the displacement, the
real parts
of the CFM roots correspond to the position of the
resonance and the imaginary parts
correspond to its width and
its damping. Of course, this interpretation is just an
approximation and we can't expect it to
precisely reproduce
the position and width of the embedded sphere resonances in
every case.  As seen before, $\Delta_l$ does not depend on
the nature of the incident wave.  Therefore its roots can be
associated
with
the embedded sphere natural vibrations which
we prefer to
term
pseudo-modes of vibrations.

Two kinds of CFM roots exist.  Some reflect primarily
natural vibrations of the sphere.  Some others are much more
damped and reflect primarily the vibrations of the
matrix around the sphere. The latter
(later referred to as ``matrix modes'') appear as a
background in the scattering calculations.  Sphere
vibrations are narrower peaks above this background.  Of
course, this distinction becomes harder to make when the
sphere and the matrix have similar characteristics (mass
density and sound velocities).  Therefore, there are several
limitations inherent to the CFM model:
\begin{enumerate}
	\item it is an approximation (because knowing the complex roots of
	$\Delta_l$ is not enough to know its variation for real arguments)
	\item what's the relative importance of sphere and matrix modes?
	\item this approach does not provide valid wave functions which are
	required for Raman scattering spectra calculations for example
\end{enumerate}

\section{CSM}
To overcome the limitations of the CFM model, it is
necessary to use a different approach.  Instead of starting
from a formulation originating from an elastic wave
scattering problem, it seems more interesting to focus on
the specificities of optical properties of nanoparticles.
In the Raman scattering problem, the vibration eigenmodes
and eigenenergies of the system are important. Then, for a
given eigenmode, the Raman intensity is related to the
displacement inside the particle (in most usual cases, Raman
scattering from the matrix is negligible in this energy
range).  In a recent paper\cite{MurrayPRB04}, we modeled the
matrix embedded nanoparticle by a core-shell system where
the ``core'' is the nanoparticle and the ``shell'' is the
matrix.  The shell's external radius is chosen several
orders of magnitude bigger than the nanoparticle radius. The
mean squared displacement within the nanoparticle
interior ($\langle u^2 \rangle_p$) was monitored by
integrating the squared
displacement field over the interior of the core as
a function of the eigenmode energy.
This uses displacement fields that were normalized
over the entire core and matrix.  As was shown, very good
agreement is obtained between the frequency and width of
particle modes calculated with CFM and peaks in the plots of
$\langle u^2 \rangle_p$ as a function of the mode frequency.
The validity of the CFM model is challenged when the
nanoparticle and the matrix are made of materials having
similar mass densities and sound velocities, for example for
a silicon nanoparticle inside silica.  In such cases, the
CFM particle modes are still in good agreement with the
peaks in the $\langle u^2 \rangle_p$ plots, but a
significant background is seen below these peaks.

\section{Conclusion}
Depending on the precision required for a given task, there
are several ways one can approach
calculations of natural vibrations of a free or embedded elastic
sphere.  If the sphere is free, then the Lamb model
suffices.  When the
sphere is not free but only the frequency
of the natural vibrations matters, the Lamb model
(free sphere model (FSM)) or the bound sphere model\cite{MurrayPRB04}
(BSM) will suffice in most cases as a workable approximation.
When the width of the resonance
matters the CFM
is an accurate tool.  CFM also fills in a few special cases
not addressed by FSM, such as the existence of matrix modes.
Finally, for situations
such as the calculation of the absolute intensity or
shape of low-frequency Raman spectra, models such as the CSM
are necessary.

\end{document}